# Pattern Recognition for the Electronic Phase of Bismuth Antimony Thin Films

Shuang Tang[1,*], Lucy Dow[1], Emmanuel Ojukwu[1]

[1] *College of Engineering, State University of New York Polytechnic Institute, Albany/Utica, NY, 12203/13502, USA*
*** E-mail**: *tangs1@sunypoly.edu*

## Abstract

There are many applications involving the use of bismuth antimony thin films. However, due to the low crystalline symmetry and strong coupling between the electronic band edges, it has always been challenging to infer the electronic phase of such a material. Fortunately, with the development of pattern recognition technology, scientists can build many black-box tools for predicting various materials properties. In this present work, we have developed several pattern recognition tools to predict the electronic phase of a bismuth antimony thin film. The support vector machine, the decision tree, and the artificial neural network are used to achieve a prediction accuracy of ~90%, ~95% and ~100%, respectively.

## 1. Introduction

Bismuth antimony thin films have been attracting intensive research interests due to its broad applications in infrared detection [1,2], supercooling [3], power generation [4], and Dirac fermion construction [5]. Bulk bismuth has a rhombohedral lattice with a $R\bar{3}m$ symmetry and can be alloyed with a broad range of antimony composition [6]. Near the band edges, there are six *H*-point, three *L*-point, and one *T*-point hole pockets, as well as three *L*-point electron pockets, which are participating in the electronic transport. The antimony composition can change the overall interatomic distances and interactions, and hence result in changes of band edge positions of the 13 carrier pockets. This leads to a great richness of electronic phases that bismuth antimony alloys can form into by varying the composition. When bismuth antimony alloys are fabricated in the form of thin films, the film thickness and the film growth orientation will further increase the complexity of its electronic band structure [7]. Due to the strong coupling between the conduction and the valence bands, the prediction of band edge positions after quantum confinement becomes especially challenging.

Tang et al. [8] have developed an iterative approach to calculate the relative positions of the 13 carrier pockets in the band structure, as a function of the antimony composition, the film thickness, and the film growth orientation with respect to the Cartesian system formed by the Bisectrix-Binary-Trigonal crystallographic directions as shown in Figure 1 (a). With different values of antimony composition, film thickness, film growth orientation are given, the electronic phase can be semimetal, direct-semiconducting or indirect-semiconducting. Further, the position of the bottom of the conduction band or the top of the valence band can be located at an *L*-point, an

*H*-point, or a *T*-point. However, since this methodology requires the operator to understand the full details of the crystallography under different asymmetric conditions, the application is still not easy.

In the recent decade, with the development of machine learning and artificial intelligence technology, researchers have used various pattern recognition algorithms [9-11], to improve the efficacy and the cost of materials property prediction, including the thermal stability [12-14], the phase changing temperature [15, 16], the mechanical strength [17, 18], the band structure [19, 20], and the band gaps [21, 22]. Owolabi et al. have used the support vector machine to generate the crystal lattice parameters of pseudo-cubic/cubic perovskites [23]. Random forest machines are used to predict the dielectric constants of over 1200 metal oxides by Takahashi et al [24]. The electronic band property of curve focal plane arrays of $InAs_{1-x}Sb_x$ alloys are described using the Gaussian process regression machine [25]. Na et al. have pointed out that a tuplewise graph artificial neural network can be used to accurately predicted the band gap of 45835 different materials [26]. Wu et al. have combined the linear Lasso, the Ridge regression, the random forest, the support vector regression, the Gaussian process regression and the artificial neural network algorithms into their pattern recognition to predict the band gaps of 3896 inorganic materials [27]. The users of these tools do not have to know all the detailed information on materials structure and solid-state physics before they can obtain a materials property with an adequate accuracy. However, for the low-symmetry system of bismuth antimony thin films, there has not been such a convenient pattern recognition method developed to predict the electronic phase.

In this paper, we will use the traditional classification methods in pattern recognition, including the supporting vector machine and the decision tree method to develop a tool for electronic phase prediction of bismuth antimony thin films. The relation of prediction accuracy and the size of training pool will be discussed. Further, we will use the artificial neural network method to achieve a better accuracy level compared to the traditional methods, by taking advantage of the intrinsic nonlinearity feature of the neuron functions.

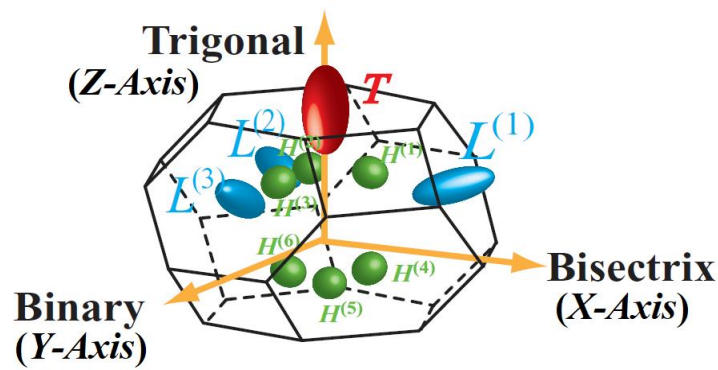

**FIGURE 1:** The three *L*-point electron pockets and the six *H*-point, three *L*-point and one *T*-point hole pockets in the Cartesian frame formed by the bisectrix, the binary and the trigonal crystalline axes of bismuth, antimony, and their alloys.

## 2. Methodology

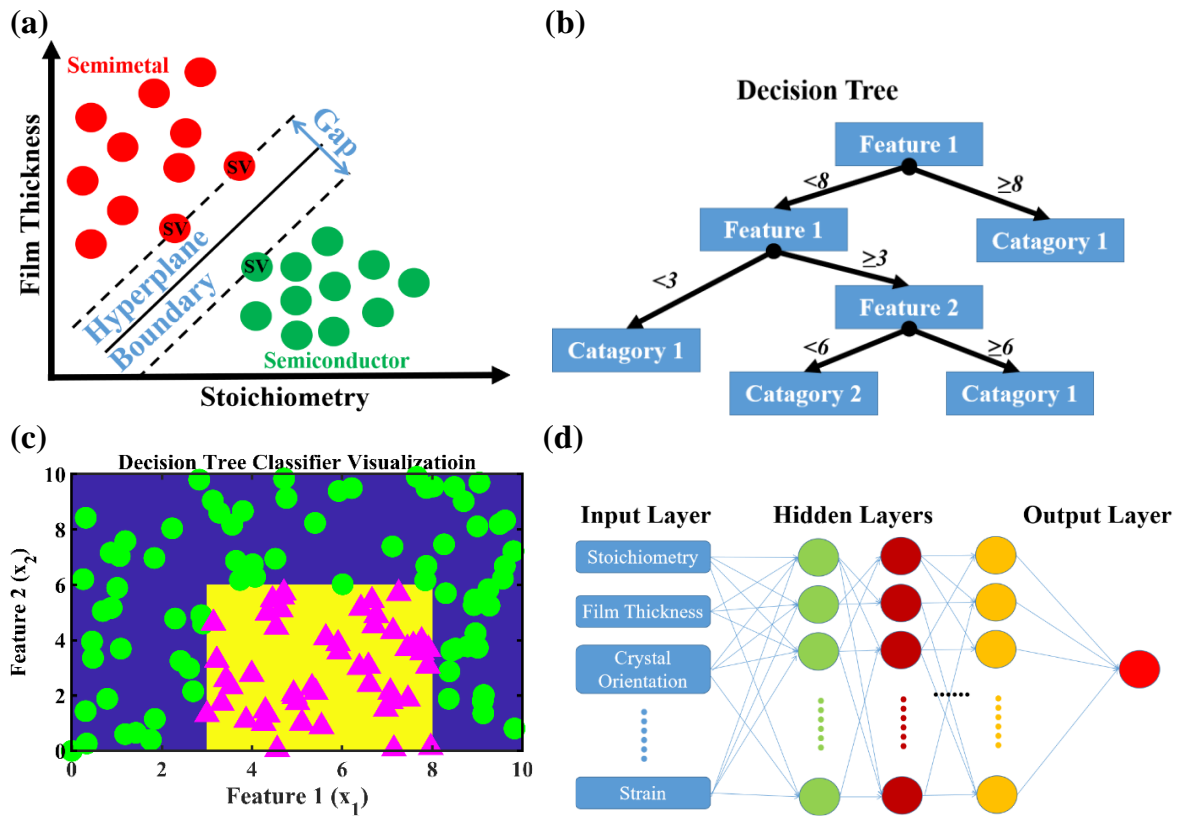

**FIGURE 2:** Illustration of common algorithms for pattern recognitions. **(a)** In a support vector machine, a hyperplane is used to approximate the boundary between each two categories in the feature space, which is obtained by maximizing the gap between the adjacent two categories. The support vectors (marked by "SV") on each side of the boundary need to be orthogonal to the hyperplane. **(b)** In a decision tree algorithm, each parent node splits the data to its child-nodes based on its own splitting rule. The recursive splitting process ends when a result or a probability distribution of results is obtained. **(c)** The boundary between two categories approximated by a decision tree algorithm can be piecewise and steplike in the feature space. In this case, the pink triangles and the green dots are representing two different categories of data, and their region are approximated by the yellow and blue colors by the decision tree algorithm. **(d)** In an artificial neural network, the feature vector is passed to the artificial neural network in the input layer. The information is then passed to one or multiple hidden layers for processing through the edges. The last layer is the output layer that gives a result or a probability distribution of the results.

## 2.1 Support Vector Machine

The support vector machine is one of the simplest and most robust supervised learning algorithms in classification applications [28, 29]. As a non-probabilistic linear classifier, a support vector machine maps the training data into the feature space and find the hyperplanes that can separate the categories of data points into their own regions.

When applied to a new sample, the data point is mapped into the feature space and its category is determined by the region that this point falls into with respect to the hyperplanes. There are many methods that can let us build these hyperplanes as the category-boundaries to separate the different categories of data points. The support vector machine chooses these hyperplane boundaries by selecting a small number of data points in the training pool from each category that can be used to build the largest gap between each two categories, which are called the support feature vectors. The category-boundary is then the hyperplane that all the support feature vectors on each side are orthogonal to, as illustrated in Figure 2 (a).

**2.2 Decision Tree Classification Method**

A decision tree is a tree architecture with internal nodes and leaf-nodes [30, 31]. Each internal node has an input feature. Each arc from one node can either lead to a subordinate decision node on a different input feature or suggest a possible result for the target, as illustrated in Figure 2 (b). The leaf nodes are either suggesting an explicit classification result or a probability distribution of different results. The tree architecture can be built by a recursive splitting rule starting from the root node and ends at the leaf nodes. At each node, a splitting rule is set based on the features of the classification. The recursive splitting continues from a mother node to the child nodes until the subset at a node has all the same values of the target variable, or until the splitting does not add values to the predictions anymore. The decision rule of each node is updated during the learning process using the training pool. When we visualize the classification, unlike the support vector machine that uses the hyperplane to serve as the category boundary, the decision tree is using multiple piecewise steplike hypersurfaces in the high-dimensional feature space to function as the separating boundaries, as illustrated in Figure 2 (c).

## 2.3 Artificial Neural Network

An artificial neural network (ANNs) is a machine learning system that mimics the brain structure of animals [32-34], which is formed by connecting layers of artificial neuron that are modeling the signal processing function of the biological synapses. When a neuron receives information from the neurons connected to it by edges within the previous layer, all the input signals are weighed, summed, and then processed by a simple nonlinear function. The neuron will then pass this processed output signal to the neurons connected to itself in the next layer. The weight associated with each edge and neuron are updated during a training process, to strengthen or weaken a synaptic connection. Besides, a threshold may exist for a neuron to stop the weak signals from passing to the next layer. Usually, only the information in the first layer (input layer) and last layer (output layer) has explicit physical meanings. All other intermediate layers are working like a black box and are, hence, called the hidden layers, as illustrated in Figure 2 (d). Depending on the application and the complexity, one or multiple hidden layers may be constructed in the neural network architecture.

## 3. Results and Discussions

To investigate how the complexity of the feature space affect the efficacy of the prediction, we will study the pattern recognition methods from a simple case to a more complex case. First, we will restrict our study in the bismuth antimony thin films that are only grown along the trigonal axis. The training pool and the testing data are chosen from the previous works [6, 35-37], with different antimony compositions and film thicknesses as illustrated in Figure 3 (a). We then build a classifier using the support vector machine, of which the classification accuracy is tested on 100 new samples. Figure 3 (b) exhibits how the classification accuracy is evolving with the training pool size. The red solid curve represents the average accuracy, and the gray area represents the uncertainty of the accuracy resulting from the choice of different training and testing samples. It is seen that the average accuracy can achieve ~90% when the training pool size is in the order of ~$10^4$. Upon enlarged training pool, the accuracy can be improved but not significantly. The same tendency exists for the uncertainty of accuracy, which is also improving with enlarged training pool, but not significantly. This can be explained by the hyperplane-building process in the linear support vector machine, which only provides a linear approximation for the real boundary between each two categories. This intrinsic drawback may be fixed by switching into a nonlinear kernel approach [38].

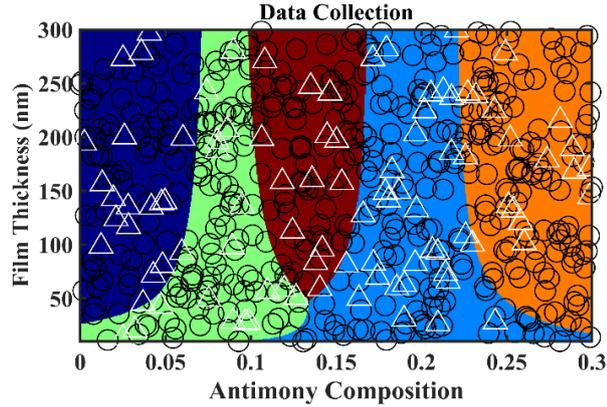

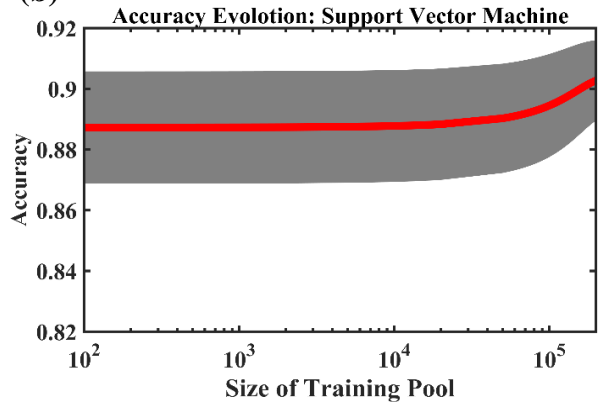

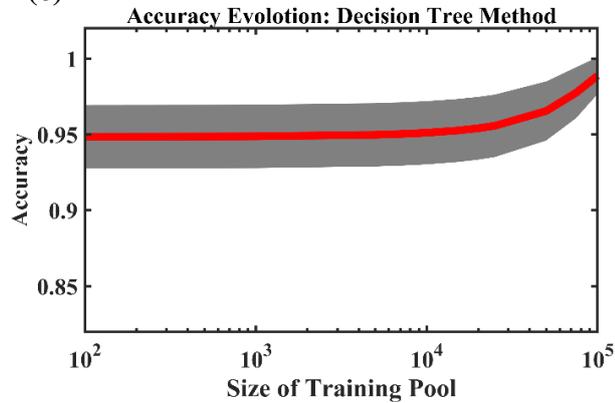

**FIGURE 3:** Electronic phase prediction for trigonal oriented bismuth antimony thin films using the support vector machine and the decision tree architecture. **(a)** Illustration of the data in the training and the test pools collected from previous works [6, 35-37] are marked, respectively, by black circles and white triangles in the map as a function of antimony composition and film thickness. The sapphire, green, brown, blue and orange colors from left to right are marking the phase regions of (1) semimetal with *L*-point electrons and *T*-point holes, (2) indirect-gap semiconductor with *L*-point electrons and *T*-point holes, (3) direct-gap semiconductor with *L*-point electrons and holes, (4) indirect-gap semiconductor with *L*-point electrons and *H*-point holes, and (5) semimetal with *L*-point electrons and *H*-point holes, respectively. The average accuracy of classification (red curve) and its uncertainty (gray area) for **(b)** the support vector machine and **(c)** the decision tree architecture is exhibited as a function of the training pool size. The uncertainty here is measured by half of the standard deviation both above and below the red curve.

We then studied the efficacy of the prediction machine built upon the decision tree architecture. The tendency of classification accuracy as a function of the training pool size is exhibited in Figure 3 (c). As expected, both the accuracy and the certainty are improving with increased size training pool. However, the accuracy can achieve ~95% when the size is in the order of ~$10^4$, which implies a much better performance than the support vector machine. Further, the shrinking of the gray area is also suggesting that the uncertainty is diminishing more obviously than the observed trend in Figure 3 (b). This can be explained by the flexibility of category-boundary shape in the feature space built by the decision tree algorithm compared to the hyperplane built by the support vector machine, which is no longer restricted to be linear. Instead, it can be a piecewise and steplike interface in the high dimensions, as we previously illustrated in Figure 2 (c).

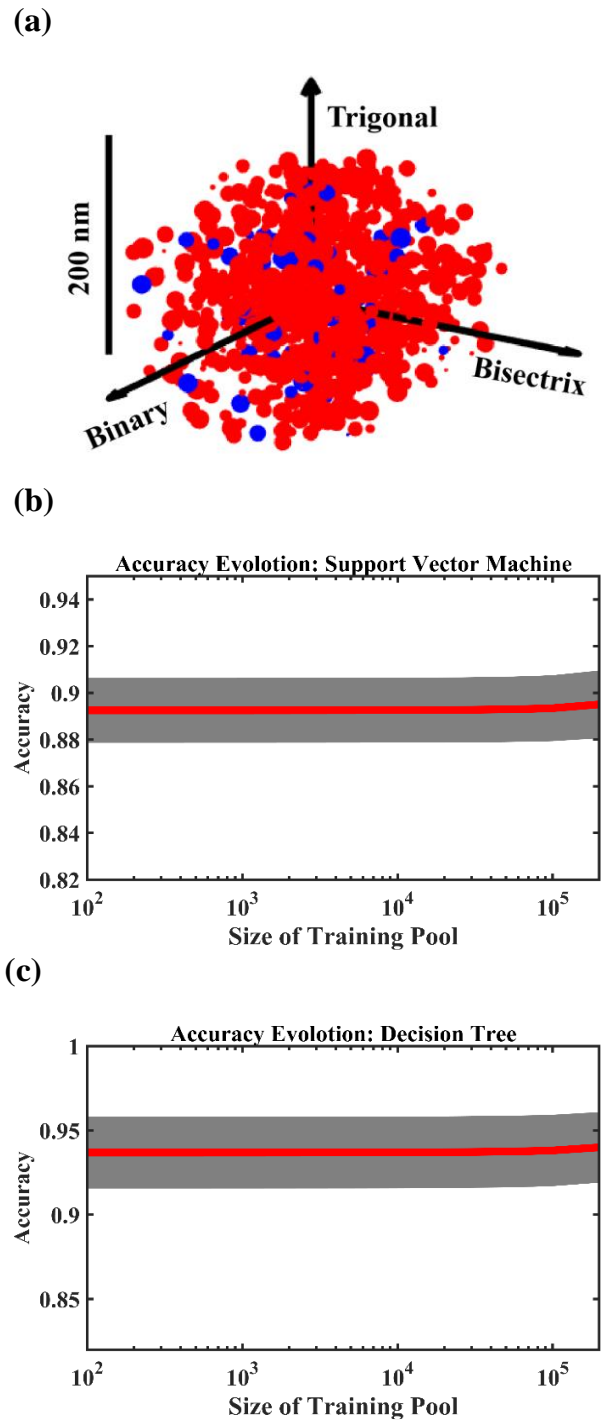

**FIGURE 4:** Electronic phase prediction for bismuth antimony thin films of a general growth orientation, using the support vector machine and the decision tree architecture. **(a)** Illustration of the data in the training and the test pools from previous works [6, 35-37] are marked, separately, by red and blue dots in the 3D space. The magnitude and the direction of the vector connecting the origin and a specific data point represent the film thickness and the film growth orientation, respectively. The size of the dot stands for the antimony composition. The average accuracy of classification (red curve) and its uncertainty (gray area) for **(b)** the support vector machine and **(c)** the decision tree architecture are exhibited as a function of the training pool size.

The efficacy of these pattern recognition tools for the trigonal direction grown bismuth antimony films is motivating us to further apply them in the general scenario of an arbitrary film growth direction. In Figure 4 (a), we are illustrating our data pool as dots in the high-dimensional feature space: the distance of each data point to the frame origin is representing the film thickness; the vector that starts from the origin and ends at the data point is representing the growth direction with respect to the bisectrix-binary-trigonal Cartesian frame; the size of each dot is representing the antimony composition. We then use both the support vector machine and the decision tree to build a pattern recognition tool for the electronic phase prediction, respectively. Figure 4 (b) and (c) are exhibiting how the classification accuracy evolves with training pool size using the linear support vector machine and the decision tree architecture, respectively. Due to the increment of complexity and the nonlinearity, the advantage of the decision tree architecture over the support vector machine becomes more obvious, compared to Figure 3 (b) and (c). The accuracy of prediction using the simple linear support vector machine seems saturated at ~89%. The enlargement of training pool is only reducing the uncertainty, but not necessarily the error rate in Figure 4 (b). On the contrary, the classification accuracy reaches a value of as high as ~95% at a very early stage and keeps improving with the training pool size, as well as the uncertainty, as shown in Figure 4 (c).

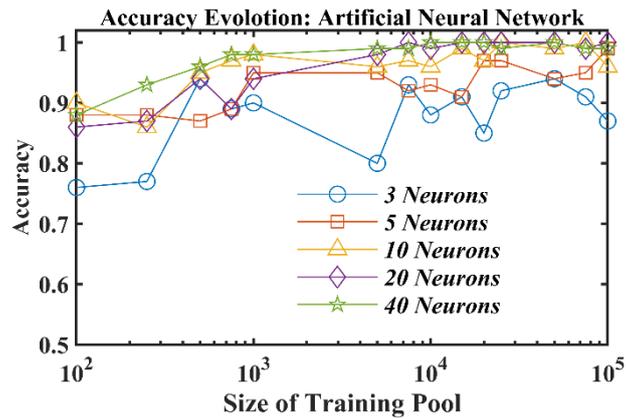

**FIGURE 5:** Comparison between the evolution of prediction accuracy with increased training pool size, when 3, 5, 10, 20 and 40 neurons are built in the hidden layer. Since our test pool is independent from the training pool, no obvious sign of overfitting is suggested here yet.

From the above comparison between the support vector machine and the decision tree architecture, we can see that the shape of category-boundaries in our feature space have non-negligible nonlinearity. Therefore, we now consider using the artificial neural network, of which the processing function in each neuron is intrinsically nonlinear. Without loss of generality, we have employed a two-layer forward artificial neural network architecture. Figure 5 (a) and (b) are showing the classification accuracy as a function of the training pool size, when 5, 10, 20, 100, and 400 neurons are used in the hidden layer, respectively. As expected, the classification accuracy is noticeably higher than the previous two algorithms, attributing to the nonlinear processing function in each neuron. However, it is more interesting to see that the increase of neurons in the hidden layer only helps in improving the classification accuracy, but does not lead to an overfitting problem, which is commonly seen in other pattern recognition problems though. This may be attributed to the fact that only one hidden layer exists in the network, so as long as the number of neurons in is larger than the dimension of the feature space, no major difference will occur upon adding or removing neurons to or from the only hidden layer.

**4. Conclusions**

The bismuth antimony alloyed thin film materials are important for many applications. However, due to the low symmetry associated with the rhombohedral crystal structure, and the strong coupling between the conduction and valence bands near their edges, the electronic phase predicting of such materials is challenging. We have developed the pattern recognition tools by treating the phase prediction as a classification problem. The traditional linear support vector machine method can give a classification accuracy of ~90% by approximating the category-boundaries using hyperplanes, the decision tree architecture can easily achieve a classification accuracy of ~95% by approximating the category-boundaries using piecewise steplike hypersurfaces, and a simple two-layer forward artificial neural network can achieve ~100% accuracy due to the incorporation of nonlinearity in the algorithm. It is also interesting to find out that the increase of neurons in the hidden layer of the network does not necessarily lead to overfitting as commonly seen in other problems. We expect this new methodology to be used not only in bismuth antimony alloys, but also in other materials systems with low degrees of symmetry and high values of entropy.

**Acknowledge**

The author(s) acknowledge the Center for Computational Innovations at the Rensselaer Polytechnic Institute for providing the AIMOS supercomputer for our research and student training.

## Author Contributions

S.T. designed and performed the research, analyzed the data, and wrote the paper. L. D. and E. O. helped in organizing and preparing of the data analysis.

## Conflict of interest

There are no conflicts to declare.

## Reference

1. I. I. Kerner, *Journal of Thermoelectricity,* 2007, 57-62.
2. J. Toudert *et al.*, *The Journal of Physical Chemistry C,* 2017,**121**, 3511-3521.
3. A. R. M. Siddique, K. Venkateshwar, S. Mahmud, B. Van Heyst, *Energy Conversion and Management,* 2020,**222**, 113245.
4. A. Nikolaeva *et al.*, *Low Temperature Physics,* 2018,**44**, 780-785.
5. S. Tang, M. S. Dresselhaus, *Nanoscale,* 2012,**4**, 7786-7790.
6. S. Tang, M. S. Dresselhaus, *Journal of Materials Chemistry C,* 2014,**2**, 4710-4726.
7. B. Lenoir, M. Cassart, J.-P. Michenaud, H. Scherrer, S. Scherrer, *Journal of Physics and Chemistry of Solids,* 1996,**57**, 89-99.
8. S. Tang, M. S. Dresselhaus, *Nano letters,* 2012,**12**, 2021-2026.
9. A. P. Bartók *et al.*, *Science advances,* 2017,**3**, e1701816.
10. Y. Zhang, C. Ling, *Npj Computational Materials,* 2018,**4**, 1-8.
11. Z. He, M. Yang, L. Wang, E. Bao, H. Zhang, *Engineered Science,* 2021,**15**, 47-56.
12. H. Zhang *et al.*, *ES Energy & Environment,* 2018,**2**, 1-8.

13. Y. Zhuo, A. M. Tehrani, A. O. Oliynyk, A. C. Duke, J. Brgoch, *Nature communications,* 2018,**9**, 1-10.

14. B. Meredig. (ACS Publications, 2019), vol. 31, pp. 9579-9581.

15. N. Qu *et al.*, *Ceramics International,* 2019,**45**, 18551-18555.

16. R. Jinnouchi, F. Karsai, G. Kresse, *Physical Review B,* 2019,**100**, 014105.

17. H. Zhang *et al.*, *Acta Materialia,* 2020,**200**, 803-810.

18. J. Li *et al.*, *Communications Materials,* 2020,**1**, 1-10.

19. Z. Wang *et al.*, *npj Computational Materials,* 2021,**7**, 1-10.

20. Z. Shi *et al.*, *Proceedings of the National Academy of Sciences,* 2019,**116**, 4117-4122.

21. A. C. Rajan *et al.*, *Chemistry of Materials,* 2018,**30**, 4031-4038.

22. Y. Huang *et al.*, *Journal of Materials Chemistry C,* 2019,**7**, 3238-3245.

23. T. O. Owolabi, *Journal of Applied Physics,* 2020,**127**, 245107.

24. A. Takahashi, Y. Kumagai, J. Miyamoto, Y. Mochizuki, F. Oba, *Physical Review Materials,* 2020,**4**, 103801.

25. A. Kyrtsos *et al.*, *Physical Review Applied,* 2021,**15**, 064008.

26. G. S. Na, S. Jang, Y.-L. Lee, H. Chang, *The Journal of Physical Chemistry A,* 2020,**124**, 10616-10623.

27. L. Wu, Y. Xiao, M. Ghosh, Q. Zhou, Q. Hao, *ES Materials & Manufacturing,* 2020,

28. T. Zhan, L. Fang, Y. Xu, *Scientific reports,* 2017,**7**, 1-9.

29. S. Chibani, F.-X. Coudert, *APL Materials,* 2020,**8**, 080701.

30. R. E. Goodall, A. A. Lee, *Nature communications,* 2020,**11**, 1-9.

31. T. Wang, C. Zhang, H. Snoussi, G. Zhang, *Advanced Functional Materials,* 2020,**30**, 1906041.

32. J. Allotey, K. T. Butler, J. Thiyagalingam, *The Journal of Chemical Physics,* 2021,**155**, 174116.

33. C. W. Park, C. Wolverton, *Physical Review Materials,* 2020,**4**, 063801.

34. W. A. Saidi, W. Shadid, I. E. Castelli, *npj Computational Materials,* 2020,**6**, 1-7.

35. S. Tang, M. S. Dresselhaus, *Physical Review B,* 2012,**86**, 075436.

36. A. D. Liao *et al.*, *Applied Physics Letters,* 2014,**105**, 063114.

37. S. Tang, M. S. Dresselhaus, *Physical Review B,* 2014,**89**, 045424.

38. C. K. I. Williams. (Taylor & Francis, 2003).

**Table of Contents Entry:**

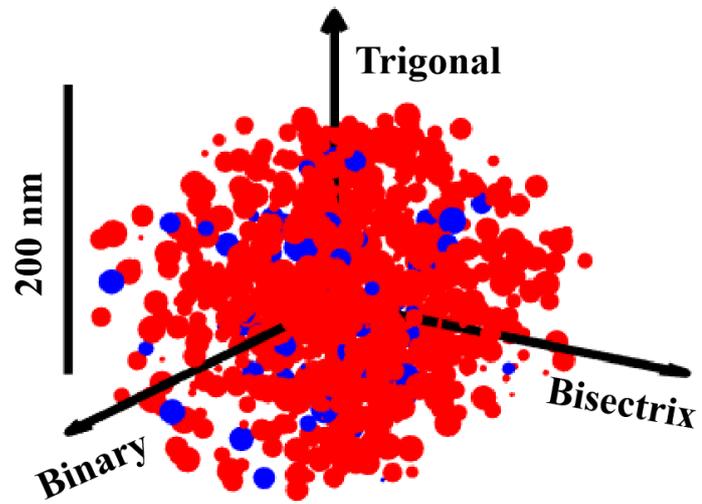

**20-word summary:**

Traditional and artificial neural network pattern recognition methods are used to predict the electronic phase of bismuth antimony thin films.